# A Model for Web Page Usage Mining Based on Segmentation

K.S.Kuppusamy[#1], G.Aghila[#2]

[#]*Department of Computer Science, School Of Engineering and Technology*
*Pondicherry University, Pondicherry, India*

*Abstract*— **The web page usage mining plays a vital role in enriching the page's content and structure based on the feedbacks received from the user's interactions with the page. This paper proposes a model for micro-managing the tracking activities by fine-tuning the mining from the page level to the segment level. The proposed model enables the web-master to identify the segments which receives more focus from users comparing with others. The segment level analytics of user actions provides an important metric to analyse the factors which facilitate the increase in traffic for the page. The empirical validation of the model is performed through prototype implementation.**

## I. INTRODUCTION

The prolific growth of World Wide Web has posed certain interesting challenges to the web content providers. These challenges include making the user to spend sufficient time on a page so that the information provided in that page would be consumed by the user. In the case of e-commerce sites this issue becomes more crucial because of the nature and intent of such sites.

Tracking the visitor activities on a page is an important source of information to enrich the site's performance in terms of content and structure. The tracking of the visitor activity need to be performed in such a manner that the web master gets the required information and the visitor's privacy is also not lost.

The user action monitoring at the page level would restrict the web master from knowing the performance metrics of internal components of the page. This research work is aimed towards solving this problem by proposing a novel approach to handle the intra-page performance. The objectives of this research work are as listed below:

- Proposing model for web usage mining at the segment level rather than at the page level.
- Validating the model empirically by conducting experiments and analysing the segment level mined data.

The remainder of this paper is organized as follows: Section II would highlight the related works carried out in this domain. Section III is about the model and the proposed algorithms. The section IV would discuss about the experiments and result analysis. Section V is on Conclusions and future directions for this research work.

## II. RELATED WORKS

The proposed model incorporates the following active research domains as listed below:
- Web Page Segmentation
- Web Usage Mining

### A. Web Page Segmentation

Web page segmentation is an active research topic in the information retrieval domain in which a wide range of experiments are conducted. Web page segmentation is the process of dividing a web page into smaller units based on various criteria. The following are four basic types of web page segmentation method:

- Fixed length page segmentation

- DOM based page segmentation

- Vision based page segmentation

- Combined / Hybrid method

A comparative study among all these four types of segmentation is illustrated in [1]. Each of above mentioned segmentation methods have been studied in detail in the literature. Fixed length page segmentation is simple and less complex in terms of implementation but the major problem with this approach is that it doesn't consider any semantics of the page while segmenting.

In DOM base page segmentation, the HTML tag tree's Document Object Model would be used while segmenting. An arbitrary passages based approach is given in [2]. Vision based page segmentation (VIPS) is in parallel lines with the way, humans views a page. VIPS [3] is a popular segmentation algorithm which segments a page based on various visual features.

Apart from the above mentioned segmentation methods a few novel approaches have been evolved during the last few years. An image processing based segmentation approach is illustrated in [4].

The segmentation process based on text density of the contents is explained in [5]. The graph theory based approach to segmentation is presented in [6].

### B. Web Usage Mining

The usage mining for web pages is studied in detail by various researchers. Three important stages of web usage mining process are data preparation, pattern discovery and pattern analysis [7].

The user interaction data can be gathered at the browser level itself through browser cache [8]. The data can be gathered through proxy server approach also [9]. The Browser agents [10] can be used to gather the usage data. The search logs [11] are also another effective way to gather the usage data.

The web logs [12] can provide crucial usage data which can be used for the personalization purpose as well. Gathering of data can be performed either explicitly or implicitly. The comparative analysis of both these types is given in [13].

In the implicit feedback scenario, various studies have [14], [15] proved that mouse activities of the user on a page can provide critical usage data.

The mouse movements serve as an important metric for user's eye tracking as well. The correlation between mouse movements and eye tracking is explored in [16], [17], [18].

In the above cited studies the analysis is at the level of a page where as in the proposed model it is fine-tuned to the segment level.

### III. THE PROPOSED MODEL

#### A. The Mathematical Model

The proposed model for web usage mining based on segmentation is illustrated in this section.

The page that the user is visiting is represented as $\Psi$ as shown (1).

$$\Psi = \left\{ \eta_1, \eta_2, \eta_3 \ldots \eta_n \right\} \qquad (1)$$

In (1) $\eta_i$ represent the segments in the page. The user visiting the page is represented as $\Gamma$. The user session on a page is as shown in (2).

$$\Gamma_{sess} = \left\{ \Psi \quad \delta \right\} \qquad (2)$$

In (2) $\Gamma_{sess}$ denotes the session of a user. It consists of two components. The first component denotes the page $\Psi$. The second component $\delta$ denotes the time spent by the user on that page.

The components in (2) can be expanded as shown in (3)

$$\Gamma_{sess} = \begin{Bmatrix} \eta_1 & \delta_1 \\ \eta_2 & \delta_2 \\ . & . \\ \eta_n & \delta_n \end{Bmatrix} \qquad (3)$$

It is not mandatory that user would visit all the segments in the page. So the session $\Gamma_{sess}$ can be expressed as in (4).

$$\Gamma_{sess} = \begin{Bmatrix} \forall \eta_i \in \Psi = |\delta_i| & if \ |\delta_i| > \sigma \\ 0 & otherwise \end{Bmatrix} \qquad (4)$$

In (4) $\sigma$ indicates the threshold limit which specifies the minimum time that user has to stay in a segment. In order to make the segment based mining process effective each of the segments would be assigned a unique number as shown in (5).

$$\Psi = \left\{ \langle \eta_1 \quad I \rangle, \langle \eta_2 \quad I \rangle \ldots \langle \eta_n \quad I \rangle \right\} \qquad (5)$$

In (5) the variable $I$ holds the unique identification number generated by the model for the segment.

The log entry made by the model for the user in a session is as shown in (6).

$$PLog(\Gamma_{sess}) = \left\{ \Psi, enTime, exTime \right\} \qquad (6)$$

In (6) enTime indicates the entry time in to that page and exTime is the exit time from the page for the user.

The $PLog(\Gamma_{sess})$ consists of a set of $SLog(\Gamma_{sess})$ components as shown in (7).

$$SLog(\Gamma_{sess}) = \left\{ \bigcup_{i=1}^{n} \left[ (\langle \eta_i \quad I \rangle, enTime), (\langle \eta_i \quad I \rangle, exTime) \right] \right\} \qquad (7)$$

The log analyser component would sort these $SLog(\Gamma_{sess})$ in the decreasing order of time spent on each segment in the page.

$$\Omega(PLog(\Gamma_{sess})) = \left\{ \langle \eta_i \quad I \rangle \qquad \delta_{i-1} \geq \delta_i \right\} \qquad (8)$$

#### B. The Architecture

The architecture diagram of the proposed model is as shown in Fig. 1. The various components involved in it are explained below:

- Segment Builder: This component is responsible for the identification of segments in the source page.

- Segment Pool: The segment pool holds the segments identified by segment builder for easier tracking.

- User Action Listener: The user action listener performs the actual tracking of mouse activities on the segments. It measures the entry time and exit time parameters for each segment.

- Session Logger: The user action listener would communicate the actions listened to the session logger for permanent storage.

- Session Log: It holds the log created by the session logger module explained above.

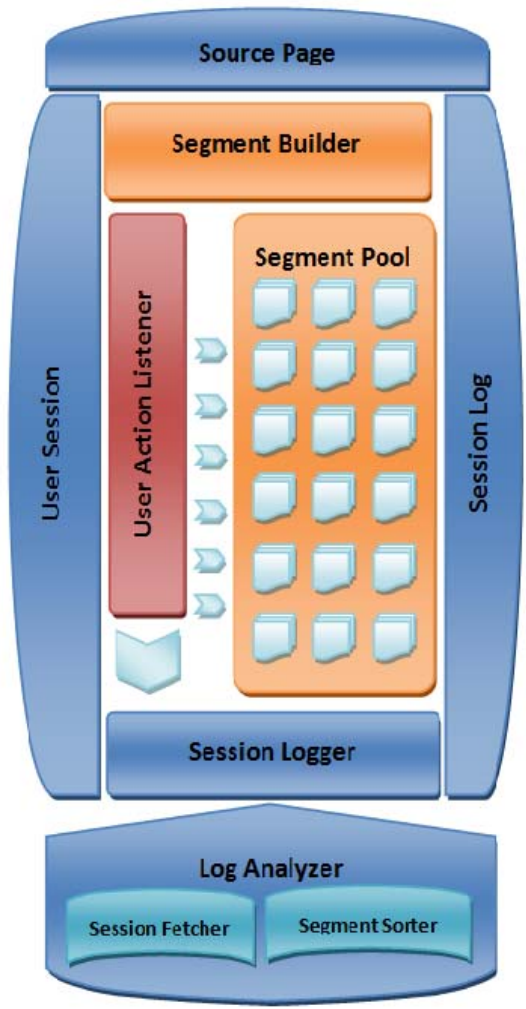

Figure 1. The Segment Based Web Usage Mining Model

- Log Analyzer: The log analyzer module's responsibility is to fetch the contents stored in the session log. The sub-component of the log analyzer i.e.

Segment Sorter would sort the segments based on the time of stay in the particular segment.

### C. The Algorithm

The algorithm for the proposed model is illustrated in this section.

The input to the algorithm is the page url P. The output of algorithm is the session log updation.

Algorithm SegmentTrack

Input   : Page P.
Output : Storage of user activity on session log.

Begin

// Segment the page
$\Psi$ = buildSegments(P)

Repeat until page is closed
begin
    locate the mouse pointer (x,y)
    seg_id = the segment for the (x,y)
    enTime = entry time into the segment;
    exTime = exit time from the segment;

    update session log (seg_id, enTime, exTime);
end

End

### IV. EXPERIMENTS AND RESULTS ANALYSIS

The experiments were conducted on the prototype implementation of the proposed model for web usage mining based on segmentation.

TABLE I USER'S INTERACTIONS WITH SEGMENTS IN A PAGE

| User Id | Segment Id | Stay (in seconds) |
|---|---|---|
| 1 | 14 | 10 |
| 1 | 23 | 25 |
| 1 | 17 | 5 |
| 1 | 6 | 4 |
| 1 | 8 | 3 |
| 2 | 45 | 22 |
| 2 | 14 | 6 |
| 2 | 22 | 6 |
| 2 | 15 | 4 |
| 2 | 14 | 3 |
| 2 | 13 | 5 |
| 3 | 17 | 10 |
| 3 | 22 | 4 |
| 3 | 8 | 4 |

| 3 | 13 | 10 |
| 3 | 22 | 15 |
| 4 | 48 | 45 |
| 4 | 14 | 12 |
| 4 | 22 | 15 |
| 4 | 15 | 35 |
| 4 | 17 | 22 |

The Table I list out the user's interactions with the segments in the page with their stay time on the page.

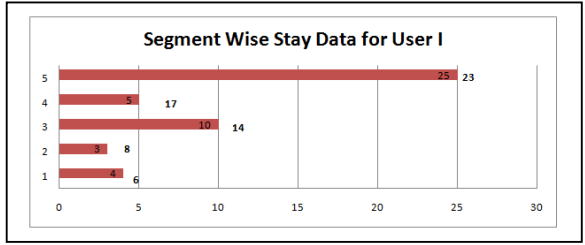

Figure 2. The Segment Interaction Data for User I

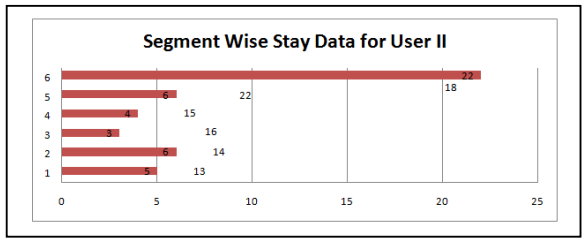

Figure 3. The Segment Interaction Data for User II

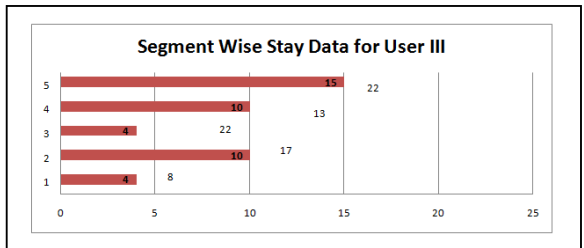

Figure 4. The Segment Interaction Data for User III

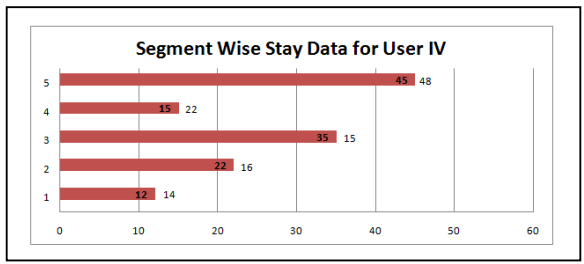

Figure 5. The Segment Interaction Data for User IV

The charts in Fig.2, Fig.3, Fig. 4, and Fig. 5 depict the user's interactions with segments in the page in-terms of their stay time.

The labels present inside the chart bars indicates the stay time and the labels out-side the chart bars indicates the segment ID.

## V. CONCLUSIONS AND FUTURE DIRECTIONS

The proposed model for web usage mining based on segmentation provides intra-page usage details. The portions of the web page which receives more focus from the users can be indentified through the proposed model.

It has been observed that different users visiting the page tend to focus on different segments on the page.

Future directions for this research work include the extension of the proposed model to handle the personalization of the page based on the mined data.

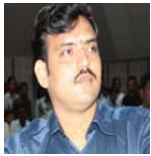

K.S.Kuppusamy is an Assistant Professor at Department of Computer Science, School of Engineering and Technology, Pondicherry University, Pondicherry, India. He has obtained his Masters degree in Computer Science and Information Technology from Madurai Kamaraj University. He is currently pursuing his Ph.D in the field of Intelligent Information Management. His research interest includes Web Search Engines, Semantic Web.

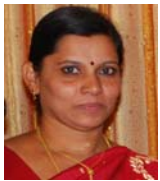

G. Aghila is a Professor at Department of Computer Science, School of Engineering and Technology, Pondicherry University, Pondicherry, India. She has got a total of 22 years of teaching experience. She has received her M.E (Computer Science and Engineering) and Ph.D. from Anna University, Chennai, India. She has published nearly 40 research papers in web crawlers, ontology based information retrieval and cheminformatics. She is currently a supervisor guiding 8 Ph.D. scholars. She was in receipt of Schrneiger award. She is an expert in ontology development. Her area of interest include Intelligent Information Management, artificial intelligence, text mining, cheminformatics and semantic web technologies